# High-power ultra-broadband supercontinuum generation in tapered multimode glass rods

Esteban Serrano, Damien Bailleul, Frédéric Désévédavy, Pierre Béjot, Grégory Gadret, Pierre Mathey, Frédéric Smektala, and Bertrand Kibler*

Laboratoire Interdisciplinaire Carnot de Bourgogne, UMR6303 CNRS-UBFC, Dijon, France
*bertrand.kibler@u-bourgogne.fr

**ABSTRACT:** Simultaneously increasing the spectral bandwidth and average output power of mid-infrared supercontinuum sources remains a major challenge for their practical application. We address this issue through experimental developments of short tapered rods made from distinct glasses (silica, tellurite and chalcogenide families) for covering distinct spectral regions by means of supercontinuum generation in the femtosecond regime. We demonstrate that ultra-broadband spectral broadenings over the entire glass transmission window can be achieved in few-cm-long segments of tapered rods by a fine adjustment of input modal excitation. As the most significant example, our simple post-processing of glass rods unlocks the high-power regime for fiber-based supercontinuum sources beyond the 10 µm waveband. By using a 5-cm-long tapered Ge-Se-Te rod pumped at 6 µm, a supercontinuum spanning from 2 to 15 µm (3 to 14 µm) with an average output power of 93 mW (170 mW) is obtained for 500-kHz (1-MHz) repetition rate. Spectral coverages from the visible to 5 µm, and from the visible to 2.7 µm, are also reported with tapered rods made of tellurite and silica glasses, respectively. Numerical simulations are used to confirm the main contribution of the fundamental mode in the ultrafast nonlinear dynamics, as well as the possible preservation of coherence features. Our study opens a new route towards the power scaling of high-repetition-rate fiber supercontinuum sources over the full molecular fingerprint region.

1. Introduction

Mid-infrared supercontinuum (SC) sources have generated many interests and found significant applications in various fields including molecular spectroscopy, remote sensing or for imaging. [1–3] The progress in this field has been possible with the deep understanding of nonlinear optical effects in optical fibers[4], in particular in the more accessible near-infrared region. In the meantime, there has been an intense activity to master materials synthesis and properties with optical fiber fabrication from suitable opto-geometrical properties, complemented by the evolution of technologies related to ultrashort pulsed lasers delivering high peak powers that are indispensable for triggering nonlinear interactions within the waveguides.[5–7] To enhance the spectral bandwidth and output power of mid-infrared SC laser sources, different types of infrared fibers, such as fluoride, tellurite and chalcogenide fibers, have been developed.[8–10] Chalcogenide (ChG) fibers, in general, feature a zero-dispersion wavelength (ZDW) situated notably distant from the near-infrared region, specifically within the range of 5-10 µm. This poses a significant hurdle given the lack of high-power and compact laser sources in this spectral range, prompting dedicated experimental endeavors to focus on fibers with shifted ZDW.[11,12] However, recent progress in tunable, high peak-power laser sources utilizing optical parametric amplifiers (OPA) and difference-frequency generation (DFG) modules, but also emerging mid-IR fiber lasers, has led to significant advancements. These advancements have substantially extended the supercontinuum and enhanced power spectral density towards the mid-infrared.[13,14]

For SC sources targeting the 2-6 µm mid-IR range, or beyond up to 10 µm, only a few studies have been carried out, in particular focusing on femtosecond pumping schemes, which can preserve the coherence features of the input laser.[6] Salem *et al.*, by pumping a step-index indium fluoride fiber with a 2 µm laser, have obtained a spectrum from 1.25 to 4.6 µm with 250-mW average power at 50 MHz repetition rate.[15] Within a two-zero dispersion step-index TeO$_2$-based fiber, Kibler *et al.* reported a spectrum spanning from 1.3 to 4.3 µm with 33-mW output power at 19 MHz repetition rate.[16] Using chalcogenide fibers, Kibler et al. also demonstrated the generation of coherent supercontinuum in the 2-6 µm range with an average power of 50 mW at 25 MHz.[14] Further in the mid-IR, Petersen et al. reported a spectrum ranging from 1 to 11.5 µm with an output power of 35.4 mW using tapered microstructured fibers.[17] Hudson et al. achieved a similar result using a tapered step-index fiber with an output power of over 30 mW for a SC ranging from 1.8 to 9.5 µm.[18] Recently a work by Tiliouine *et al*. demonstrated an all-fiber format source with a SC spanning from 1.7 to 7.8 µm with 20 mW output power.[19] Nevertheless, it is worth noting that the cascaded fiber architecture has remained the most suitable and developed solution for developing compact SC sources with high power and broadband mid-IR spectrum[20], but at the expense of the coherence degradation. Until now no solution exists to combine all the advantageous properties (compactness, power scaling, coherence, simple fiber design) of the above developments to go beyond the frontier of the 10-11 µm waveband with optical fibers.[20]

In this work, we demonstrate the significant potential of short tapered multimode rods for high-power ultrabroadband SC generation on distinct glass platforms by means of a high-repetition-rate femtosecond pulse pumping scheme. Our solution reveals state-of-the-art achievements in terms of average output power and spectral coverage. Typically, we report a SC source spanning from 2 to 15 µm with a 93-mW average power at 500-kHz repetition rate in a tapered chalcogenide rod. We also confirm by numerical simulations based on the generalized nonlinear Schrödinger equation (GNLSE) for different pumping powers that the spectral broadening mainly involves nonlinear dynamics in the fundamental mode of tapered glass rods. Remarkably, the predicted first-order coherence of the largest mid-IR SC is shown to be preserved as high as the pumping laser. Finally, we discuss power scaling, future investigations, and potential limitations of the tapered glass platform for SC generation and its application.

## 2. Methods

2.1. Materials

a) Ge-Se-Te glass

We used a chalcogenide rod (i.e., single-index fiber) with 180-µm diameter, made of glass from the Ge-Se-Te ternary system. Prior studies have detailed the synthesis of Ge-Se-Te glasses.[11,14,21] Specifically, the selected composition ($Ge_{20}Se_{70}Te_{10}$) belongs to the Se-rich region of the $GeSe_4$-$GeTe_4$ pseudo-binary system, chosen to ensure compatibility with fiber technology. A $Ge_{20}Se_{70}Te_{10}$ glass preform was synthesized by means of the standard melt quenching technique[21]. Our purification process allowed to significantly reduced the absorption bands, particularly those associated with O and H bonds. The 16-mm outer diameter preform was drawn into a thin optical rod (i.e., single-index fiber) measuring 180 µm in diameter and several meters in length. We evaluated the linear losses of the Ge-Se-Te rod by using the cut-back method. Notably, we observed a minimal continuous background, measured below 2 dB/m (in the 5-9 µm spectral range). In Figure 1(a), calculated dispersion curves of the fundamental mode only are presented for rod diameters of both 180 and 40 µm. As expected, the 180-µm diameter rod exhibits dispersion properties very similar to the bulk material (see Ref.[21] for refractive index data), with a zero-dispersion wavelength (ZDW) at approximately 7.1 µm. Upon reducing the rod diameter to 40 µm, the ZDW shifts towards shorter wavelengths, reaching nearly 6.3 µm. Over the wide 2-16 µm wavelength range, even for the smallest waist diameter (40 µm), note that the guiding feature of our thin rods remains multimode.

Figure 1(d) further illustrates the nonlinear coefficient $\gamma$ associated with the calculated fundamental mode (insets depict the corresponding modal distribution at the pumping wavelength). The relationship is given by $\gamma = 2\pi n_2/(\lambda A_{\text{eff}})$, where $\lambda$ is the wavelength, $A_{\text{eff}}$ the corresponding effective (fundamental) mode area, and $n_2 = 8.2 \times 10^{-18}$ m².W$^{-1}$ is the nonlinear index of our glass.[22] This nonlinear coefficient experiences a strong increase for smaller rod diameters while the fundamental mode remains well confined in the glass rod due to the high refractive index difference. For instance, at the pumping wavelength of 6 µm, the nonlinear coefficient is 20 times higher for a 40-µm tapered rod compared to the initial 180-µm rod. Given that we will be operating within highly nonlinear propagation regimes characterized by high pulse peak power, it becomes crucial to know the critical self-focusing power for each material. This critical peak power can be estimated with the following expression[23], $P_{\text{crit}} = 1.86 \lambda_0^2/(4\pi n_0 n_2) = 0.26$ MW, where $\lambda_0 = 6$ µm is the pumping wavelength and $n_0 = 2.494$ is the refractive index at $\lambda_0$ for our Ge-Se-Te glass.

b) $TeO_2$-ZnO-$La_2O_3$ glass

As tellurite glass, we employed for the single-index fiber the following composition: $70TeO_2$-$25ZnO$-$4.25La_2O_3$-$0.75La_2F_6$, with the introduction of fluorine aimed to reduce OH contamination in the mid-infrared range. Prior works have previously outlined the synthesis and purification of glass from the $TeO_2$-ZnO-$La_2O_3$ ternary system (TZL).[24,25] Preform syntheses are conducted using the classical melt-quenching method. The entire fabrication process takes place under a controlled atmosphere of dry air, where the $H_2O$ concentration is meticulously monitored to maintain it below 0.5 ppm (vol.). The optical losses of this single-index fiber exhibit a continuous background below 0.6 dB/m from 1.8 µm to 2.8 µm and remain below 8 dB/m up to 4 µm[25]. Reference[26] provides detailed measurements of the refractive index variation of the glass with wavelength. The TZL glass rod exhibits an initial outer diameter of 160 µm. The corresponding dispersion properties for the fundamental mode are depicted in Figure 1(b) for rod diameters of both 160 and 40 µm. The ZDW is approximately 2.31 µm for the initial diameter and slightly shifts to 2.29 µm in the tapered diameter. At 2.5-µm wavelength, the index difference between the glass and the surrounding air enables the fundamental mode to be well confined at all fiber diameters. The tellurite rod also exhibits a multimode guiding behavior over the wavelength range of transmission. The wavelength-dependent nonlinear coefficient of the fundamental mode is shown in Figure 1(e). Notably, the nonlinear coefficient is substantially higher in the tapered region (here $n_2 = 3.8 \times 10^{-19}$ m².W$^{-1}$),[27] and reaches a level that is 16 times higher than that of initial diameter region at 2.5-µm wavelength. As shown in Figure 1(e), the fundamental mode confinement is highly effective for both diameters. For our TZL glass, the critical self-focusing peak power is found to be $P_{\text{crit}} = 1.24$ MW at the pumping wavelength $\lambda_0 = 2.5$ μm (with $n_0 = 1.969$).

c) SiO$_2$ glass

As a silica glass rod, we simply used a commercially-available step-index fiber (Thorlabs, FG105LCA) with a large pure silica (SiO$_2$) core with a diameter of 105 μm, surrounded by a thin cladding of fluorine-doped silica (F-doped SiO$_2$). The corresponding numerical aperture is 0.22. The optical losses of this fiber is about 0.001 dB/m between 0.8 and 1.9 μm, increasing to 0.08 dB/m up to 2.2 μm. The external diameter of the multimode fiber is 125 μm. For our calculations of the fundamental mode properties, we considered the material dispersion of pure silica glass based on the Sellmeier formula given in Ref.[28]. The refractive index difference between the core and cladding glasses was set to an approximate constant value, denoted as $\Delta n = 0.01675$ (see Ref.[29]). Figure 1(c) displays the dispersion for the initial diameter (125-μm external diameter), and for a tapered rod diameter of 25 μm. The variation in the ZDW position is less pronounced here in the case of the silica fiber. Figure 1(c) shows that the ZDW is shifted down to 1.25 μm for the 25-μm waist diameter, whereas it is located around 1.27 μm for the initial diameter. The wavelength-dependent nonlinear coefficient of the fundamental mode is shown in Figure 1(f). For 1.5 μm wavelength, the fundamental mode remains well confined for the initial and the waist diameter, as shown in the inset. This fiber is also characterized by a multimode guidance, although it is less multimode than the chalcogenide fiber or the tellurite fiber in the 1-4 μm spectral range, for the 25-μm waist diameter. The nonlinear coefficient at the pump wavelength ($\lambda_0 = 1.5$ μm) can be significantly increased up to 20 times. We again calculated the critical self-focusing peak power, as $P_{\text{crit}} = 8.86$ MW at the pump wavelength (with $n_0 = 1.445$ and $n_2 = 2.6 \times 10^{-20}$ m$^2$.W$^{-1}$)[30].

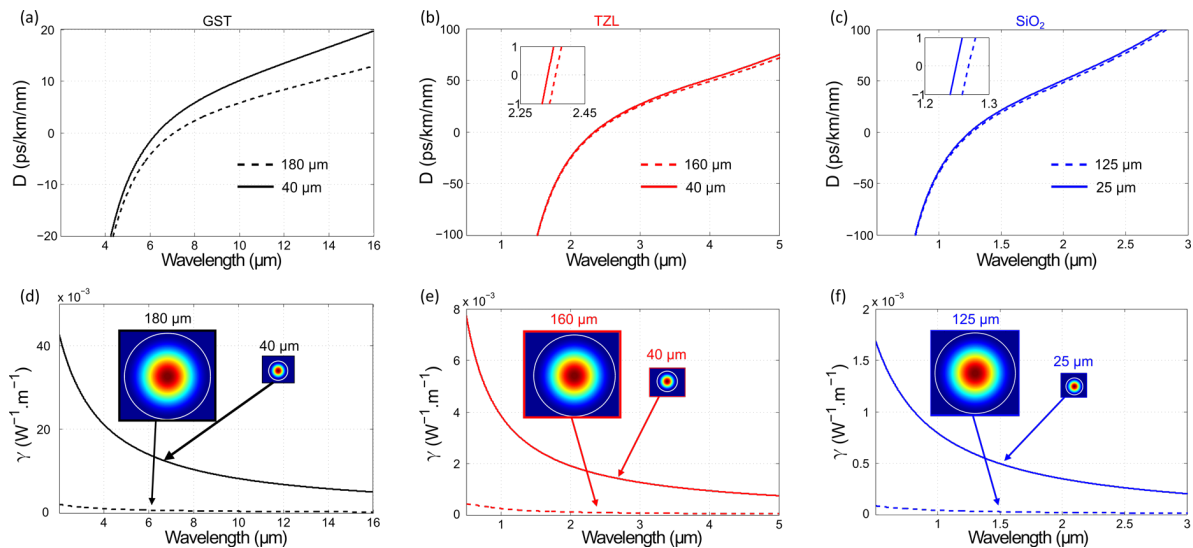

Figure 1- (a-c) Wavelength-dependent curves of dispersion $D$ of the fundamental guided mode of single-index fibers with distinct rod diameters in the cases of GST, TZL, and SiO$_2$ fibers. (d-f) Calculated wavelength-dependent curves of rod nonlinear coefficient for the same diameters.

2.2. Experimental setup used for SC generation

To produce short adiabatic biconical tapers from our initial fiber, a well-established post-processing technique was employed utilizing the VYTRAN Glass Processing Workstation (GPX-3400). This glass processing platform facilitates fusion splicing and tapering of specialty fibers, equipped with distinct filament heaters, precision stages with multi-axis control, and a high-resolution CCD imaging system. For both chalcogenide and tellurite rods, an iridium heating filament (FRAV3) is used, enabling work to be carried out at temperatures between 150 °C and 600 °C for diameters < 400 μm. For the silica fiber, a graphite filament is used to work at higher temperatures (FTAV2). In our configuration, we can achieve a tapering factor up to 1/12 onto the fiber diameter, creating a maximum 80-mm-long waist region. Through trial and iterative methods, the transition length was optimized to 12.5 mm, while the waist length was set between 15 and 25 mm. The choice of the waist diameter for each glass system involves considering various parameters. The goal is to choose a diameter that not only enhances the nonlinearity, including the shift of the ZDW, but also withstands the mechanical demands of our experimental work. By a trial-and-error method, we produced tapered rods with a waist diameter of 40-μm for the GST and TZL glasses, and a waist diameter of 25-μm for silica. The selection of the laser pump wavelength depends on both the pulse energy available and the ZDW of the glass rod under study. In the case of GST fiber, the pump is set at 6 μm (in the normal dispersion regime), while for TZL and silica rods, the laser pump is chosen in the anomalous dispersion regime (2.5 μm and 1.5 μm, respectively). Fabricating a tapered rod typically results in increased transmission losses of nearly 2-3 dB in average compared to the uniform rod, such values were obtained for multimode linear propagation. Such additional transmission losses may arise from two main contributions: (i) the non-ideal taper profile and surface defects, and (ii) the spatial filtering of higher-order modes in the waist section when multimode propagation takes place.[31]

The practical testing of our tapered rods for mid-IR SC generation was conducted by means of the experimental setup illustrated in Fig. 2. Infrared pump pulses at distinct wavelengths for SC generation were obtained using a high-power femtosecond laser (Monaco 1035, Coherent) that delivers 260-fs pulses at 1035 nm, with a repetition rate ranging from 10 kHz to 1 MHz, followed by an optical parametric amplifier (OPA) to produce a signal beam tunable from 1.35 to 2.06 µm and an idler beam tunable from 2.06 to 4.5 µm. Signal pulses (with 7 µJ maximal energy) were employed to pump our silica rods, and idler pulses (with 3 µJ maximal energy) were used to pump our tellurite rods. In order to pump our chalcogenide rods, the signal and idler beams are combined and passed through a difference frequency generation (DFG) unit to provide mid-IR pulse tunable from 4.5 to 16 µm (with 0.85 µJ maximal energy). In any case, the delivered pulse energy was adjusted to µJ-level energies by means of neutral density filters to avoid irreversible damages during coupling optimization. We then coupled the IR laser beam into the rods under investigation by means of suitable lenses (typically with long focal lengths, 40 and 125 mm for soft-glass and silica rods, respectively), chosen through an optimization process to adjust the beam diameter at the focal point. This adjustment aims to possibly match the calculated fundamental mode diameter of the glass rod. As shown in the following, the preferential excitation of the fundamental mode is crucial for optimizing SC properties in such highly multimode rods. This approach favors an adiabatic change of the mode along propagation, minimizing energy loss occurring in the higher-order modes which may become non-guided in the tapered region. Even in the ideal case of perfect fundamental mode excitation, combined with the femtosecond regime and short propagation distances, the observation of spontaneous intermodal phase-matching is possible. This phenomenon is associated with the discretized conical emission, especially when employing input peak powers in the same range as the critical self-focusing threshold.[32,33]

At the glass rod output, the SC light was collected by means of a 1-m-long hollow core fiber from Guiding Photonics with an internal diameter of 1.5 mm and internal dielectric coatings for SC spectra extending beyond 5-µm (i.e., for chalcogenide rods), otherwise a 400-µm core diameter $InF_3$ fiber was used. Subsequently, the collected SC light was sent to two different systems for spectral analysis, depending on the type of glass rod examined (see Figure 2). In the case of chalcogenide rod, the first detection method involves injecting the SC into a monochromator equipped with various gratings connected to a Mercury Cadmium Telluride (MCT) detector capable of operating up to 22 µm. Filters are incorporated to prevent high-order diffraction peaks from the gratings. For the silica or tellurite rods, the second detection method involved the use of an optical spectrum analyzer operating from 1.9 to 5.5 µm (Yokogawa, AQ6377), along with a FTIR spectrometer for wavelengths below 1.9 µm. For both chalcogenide and tellurite rods, we were not able to analyze the spatial distribution of the spectrum due to the absence of suitable camera covering the ultrabroad spectral range. We only characterized the SC spatial distribution for silica rods by imaging the output facet onto a visible/near-infrared camera. For all the glass rods, the average power of the output SC is measured using the same thermal power sensor.

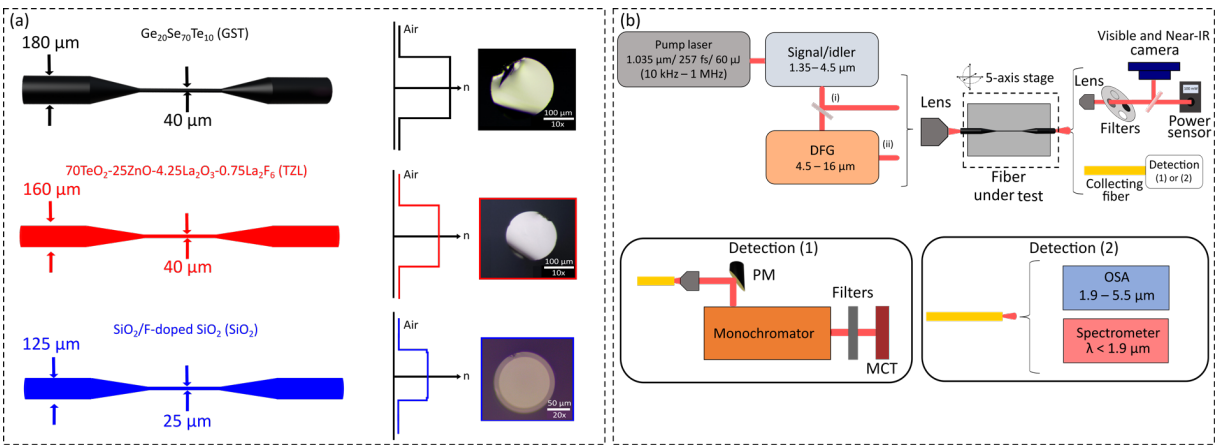

Figure 2- (a) Schematic view of optimized tapered rod for each glass (left side), and corresponding refractive index profiles and input cross-section images (right side). (b) Experimental setup used for SC generation. The laser pump used can either come from (i) the signal and idler or (ii) the DFG module. The fiber under study is placed on a V-groove mounted on a 5-axis stage. Output SC is characterized with detection (1) for GST fibers and detection (2) for TZL and $SiO_2$ fibers. A power sensor is also used to characterize the output power of the spectrum and a visible/near-IR camera for near-field imaging.

2.3. Numerical modeling

To model nonlinear pulse propagation and SC generation in our uniform or tapered glass rods, we made use of the widely employed generalized nonlinear Schrödinger equation (GNLSE).[6] This equation incorporates the full dispersion curve of the fundamental guided mode for each glass rod, accounting for both instantaneous Kerr and delayed Raman nonlinear responses, as well as the dispersion of nonlinearity, considering the frequency dependence of the effective mode area. For the Raman response function, we employed an intermediate-broadening model utilizing convolutions of Lorentzians and Gaussians. This model is adapted from spontaneous Raman scattering spectra from each glass, incorporating the estimated Raman gain coefficient referenced in literature.[22,34–36] Our simulations also include linear losses. For tapered rods, the longitudinal variations of modal properties are fully taken into account as a function of the rod geometry. Note that a most suitable modeling could be developed based on the recently derived multimode unidirectional pulse propagation equation[29,30,32,37] instead of using the simpler GNLSE that fits a single-mode propagation behavior, however, this would require the non-trivial introduction of longitudinal variations of modal properties as well as the detailed knowledge and characterization of input spatial coupling at distinct wavelengths. Furthermore, the ultra-broadband spectral window investigated here would also imply a significant computation time cost. In the following, we show that a careful control of fundamental mode excitation can simplify the modeling issues, by strongly limiting the energy expansion over higher-order modes. Over the past decade, extensive research has explored the correlation between the stability of fiber-based SC and the characteristics of input pulses, as well as fiber parameters.[38,39] The anomalous dispersion regime is recognized for promoting the interplay between coherent processes (soliton dynamics) and incoherent mechanisms (modulation instability) responsible for spectral broadening. It is widely accepted that maintaining an input pulse energy equivalent to a soliton number $N$ below 16 is a general guideline to preserve high coherence over the full SC spectrum and stability from pulse to pulse. The soliton number is defined as $N^2 = L_D/L_{NL} = \gamma P_0 T_0^2/|\beta_2|$, where $P_0$ is the pulse peak power, $T_0$ is the half-width (at 1/e-intensity point) of the input pulse, and $\beta_2$ is the group velocity dispersion at the pump wavelength. Here, $L_D = T_0^2/|\beta_2|$ represents the dispersion length, and $L_{NL} = 1/\gamma P_0$ is the nonlinear length. In contrast, fibers with all-normal dispersion allow the design of highly coherent SC with initial $N$ values reaching up to a few hundreds. Coherence properties associated with the SC spectrum were here analyzed in a manner similar to Ref.[6] Specifically, we conducted 200 simulations, each with different random input noise applied to the input pulse for each glass rod. The resulting ensemble of output SC fields enabled the calculation of the modulus of the complex degree of first-order coherence $|g_{12}^{(1)}|$ defined at each wavelength in the SC. This parameter is commonly used to characterize the shot-to-shot stability of SC sources.

## 3. Results

### 3.1. SC generation in Ge-Se-Te rods

Our initial investigation focused on SC generation in uniform and tapered Ge-Se-Te rods with the pump wavelength set to 6 µm and 250 kHz-repetition rate. The strong nonlinearity of this composition led to a significant spectral broadening spanning from 3 µm to 9 µm in the uniform rod, particularly at the highest peak power injected equal to 576 kW (twice the estimated self-focusing threshold $P_{\text{crit}}$), as shown in Fig. 3a. Experimental findings align well with numerical results based on the GNLSE, with a relatively small deviation mainly at shorter wavelengths (see Fig. 3b), likely attributed to the largely multimode nature of the glass rod and initial self-focusing occurring in the first steps of propagation. Thus, despite efforts to couple the input pulse into the fundamental mode, the possible spontaneous formation of discretized conical wave may occur, which generally involves low energy couplings (in the form of phase-matched dispersive waves) into higher-order modes at far-detuned frequencies from the pump.[33] When using the 40-µm-waist diameter tapered rod, we easily observe a considerable enhancement of mid-IR SC generation for the same distinct powers investigated (see Fig. 3c). This improvement can be attributed to the blue-shift of the zero-dispersion wavelength (ZDW) in the waist region towards the pump wavelength and more importantly to the stronger light confinement both shown in Fig. 1(a,d). For the same maximal input peak power as previously, we note that the SC now covers the spectral range from 3 µm to 12.5 µm. Remarkably, for all the SC spectra recorded, the agreement with numerical simulations has improved on both spectral edges. This may arise from the spatial filtering of higher order modes in the tapered section, which also implies the additional losses discussed in the previous section. As a guideline, it is worth mentioning that the calculated number of guided $LP_{0m}$ scalar modes strongly decreases from 69 to 15 when the initial 180-µm rod diameter is tapered down to 40 µm. The average power measurements at the taper output corroborate the above description, namely we consecutively measured 28%, 39% and 95% of the power measured at the uniform rod output for the same input peak powers (576, 231, and 45 kW). This clearly means that at low powers (below self-focusing threshold), there is almost no additional loss between tapered and uniform propagations when exciting mainly the fundamental mode. However, when strongly increasing the input power, the total transmission of the tapered rod decreases due to spatial filtering of higher-order modes in the waist section. Such higher-order modes are excited due to self-focusing occurring in the first steps of propagation and further intermodal couplings and energy exchanges may occur. As a result, we considered the measured output powers for the tapered rod as a suitable indicator of power contained in the fundamental mode for our numerical modeling based on GNLSE. Thus, 164, 90, and 44 kW were the values of peak power used for generating numerical results shown in Fig. 3d. Although we faced challenges in characterizing the modal content through near-field imaging due to the absence of an infrared camera covering the entire SC range, we can assume that higher-order mode filtering occurs in the tapered section, resulting in a less multimode output spectrum.[40]

In terms of coherence, it is interesting to observe that even at the highest peak power beyond the self-focusing threshold, we can preserve full coherence of the SC over the entire wavelength range (when considering only the fundamental mode). The main origin of this advantage can be found in the input pumping configuration occurring under the normal dispersion regime (below the ZDW). From simulations, the main nonlinear dynamics involved in the spectral broadening remain the self-phase modulation that generates a SC broad enough to extend beyond the ZDW and take advantage of the dynamics of the anomalous dispersion regime. In this case, the calculated N values in the simulation was 5 in the initial diameter and more than 30 in the 40-µm waist diameter.

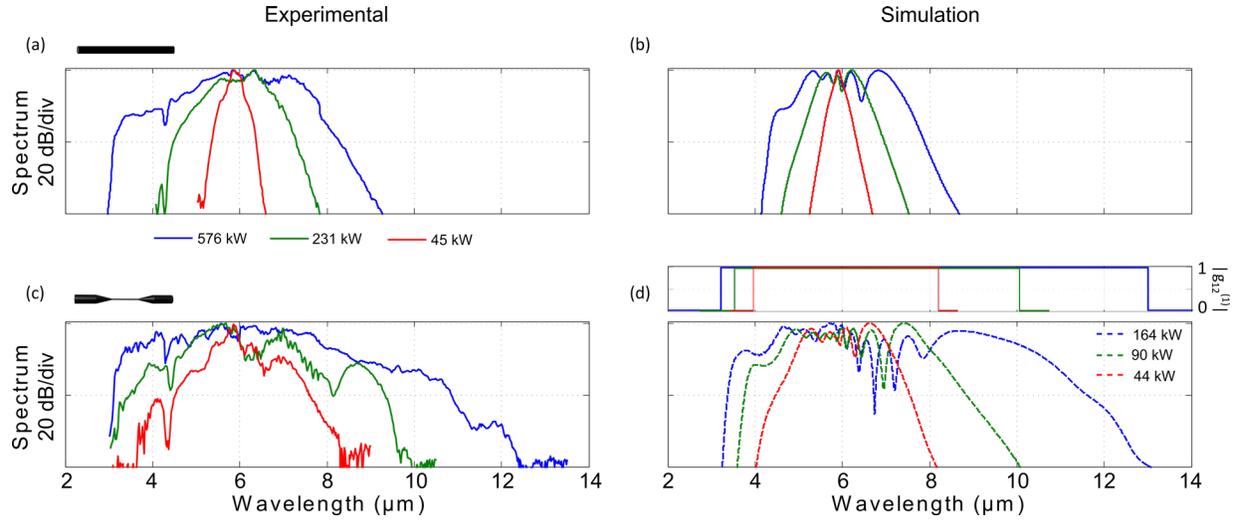

Figure 3- SC spectra obtained in 5-cm long segments of single-index Ge-Se-Te glass rods with (a-b) uniform and (c-d) tapered sections and different input peak powers. Left panels: Experimental measurements. Right panels: corresponding numerical results of SC generation obtained over 200 simulations. The calculated modulus of complex degree of first-order coherence is also depicted in (d) for the tapered rods.

### 3.2. SC generation in TeO$_2$-ZnO-La$_2$O$_3$ rods

In a second set of experiments, we investigated SC generation in our uniform and tapered TZL rods. The pump wavelength was here set to 2.5 µm and the repetition rate to 10 kHz. The uniform TZL rod exhibits anomalous dispersion at the pump wavelength (see Fig. 1b). Figure 4(a-b) shows corresponding experimental and numerical results obtained with different input peak powers for the uniform rod with 160-µm diameter. Even in this simple glass rod with moderate nonlinearity, we obtained a significant spectral broadening spanning from 0.9 µm to 3.8 µm, owing to the high peak power injected up to 2.9 MW (about twice the estimated self-focusing threshold $P_{\text{crit}}$). As previously observed for the chalcogenide rod, our simple numerical simulations well confirm the SC shape and bandwidth for increasing input powers, except on the short-wavelength edge that can be related to higher-order modes. From the simulations, the spectral broadening is primarily governed by soliton dynamics and the formation of phase-matched dispersive waves occurring below the zero-dispersion wavelength.

Once again, we confirm that using a tapered rod clearly improves the resulting SC bandwidth as well as its flatness even at lower peak powers, as shown in Fig. 4c, particularly due to the stronger modal confinement in the waist region (see Fig. 1e). The SC now covers the spectral range from 0.6 µm to 4.8 µm. At the taper output, we measured nearly 77% of the power measured at the uniform rod output for all the input peak powers (2.9, 1.9, and 0.5 MW). Corresponding peak power values of 2.29, 1.46, and 0.38 MW were used for generating numerical results shown in Fig. 4d. Numerical simulations consistently corroborate our experimental findings while still having a slight deviation at shorter wavelengths. Compared to the Ge-Se-Te taper, the total transmission is constant, but it is lower at the lowest input power (below the self-focusing threshold), thus indicating a poorer excitation of the fundamental mode. Moreover, the number of LP$_{0m}$ modes from the initial diameter to the waist diameter decreases from 109 to 27. In other words, the present TZL rod is characterized by a stronger multimode behavior, thus reducing the possible modal filtering. Unlike the chalcogenide case, here the increasing peak power significantly degrades the SC coherence in particular as soon as approaching the self-focusing threshold, since we reach input soliton number $N > 10$ and we are simultaneously pumping in the anomalous dispersion regime. Note that for the 160-µm diameter the calculated input $N$ value is 11 when P$_0$ = 2.29 MW but it grows beyond 30 for the 40-µm waist section due to a higher $\gamma$ parameter (16 times higher) with the smallest diameter.

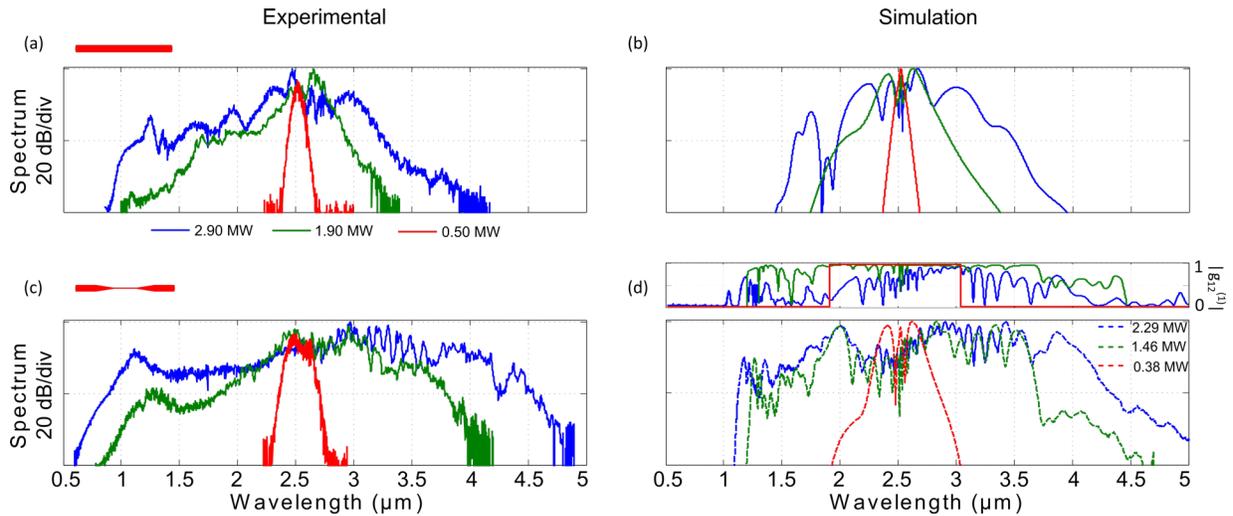

Figure 4- SC spectra obtained in 5-cm long segments of single-index TeO$_2$-ZnO-La$_2$O$_3$ glass rods with (a-b) uniform and (c-d) tapered sections and different input peak powers. Left panels: Experimental measurements. Right panels: corresponding numerical results of SC generation obtained over 200 simulations. The calculated modulus of complex degree of first-order coherence is also depicted in (d) for the tapered rods.

### 3.3. SC generation in SiO$_2$ rods

In a third set of experiments, we investigated SC generation in our uniform and tapered silica fibers with a pump wavelength set to 1.5 µm and a repetition rate of 125 kHz. For the uniform fiber, the 5-cm length is not effective for generating any supercontinuum (see Fig. 5a), even at the highest peak power of 2.65 MW, that is well below the self-focusing threshold. Again, this behavior is fully confirmed by our numerical simulations based on the GNLSE (see Fig. 5b), and it can be easily related to the low nonlinearity of silica glass. In the same way as for the other materials, with the tapered fiber, SC generation becomes possible and considerably efficient, with a spectral coverage ranging from 0.4 to around 2.7 µm, as reported in Fig. 5c. At the taper output, we consecutively measured 85%, 93% and 97% of the power measured at the uniform fiber output for the same input peak powers (2.65, 1.12, and 0.9 MW). There is almost no additional loss at moderate and low powers between tapered and uniform propagations when exciting mainly the fundamental mode. By increasing the input power, the total transmission slightly decreases and reveals some spatial filtering of higher-order modes in the waist section. The number of LP$_{0m}$ modes from the initial diameter to the waist diameter decreases from 16 to 3, thus showing the restrained multimode guidance when compared to previous glass rods investigated.

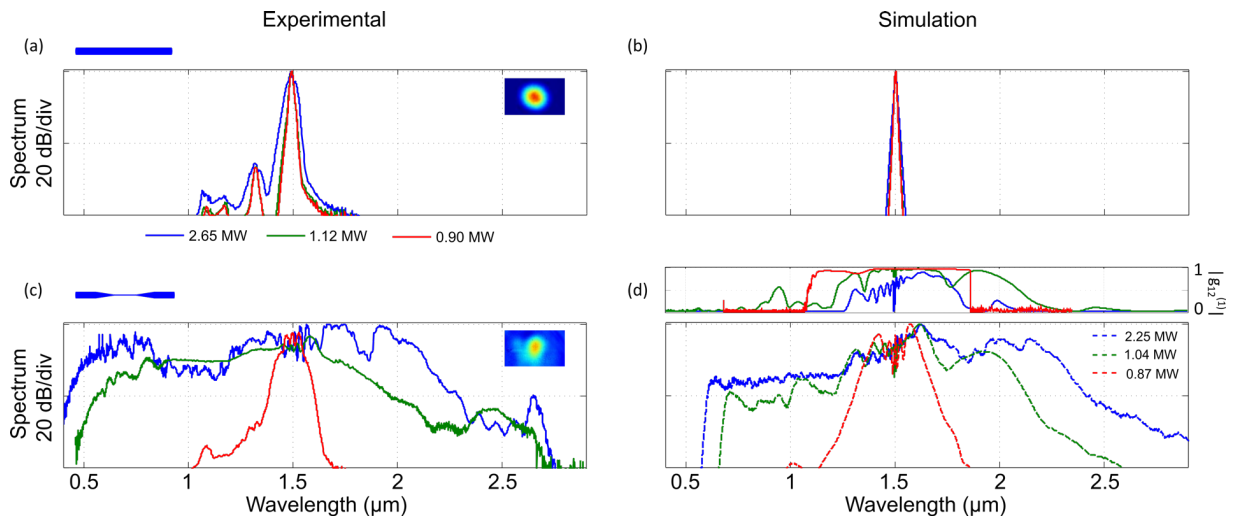

Figure 5- SC spectra obtained in 5-cm long segments of multimode step-index SiO$_2$ fibers with (a-b) uniform and (c-d) tapered sections and different input peak powers. Left panels: Experimental measurements. Right panels: corresponding numerical results of SC generation obtained over 200 simulations. The calculated modulus of complex degree of first-order coherence is also depicted in (d) for the tapered fibers.

Corresponding peak power values of 2.25, 1.04, and 0.87 MW were used for generating numerical results shown in Fig. 5d. The spectral broadening is well confirmed by numerical simulations for distinct input peak powers, except on the short wavelength edge. Here, although the high peak powers used, all values are below the critical power of self-focusing. However, for the same reasons as TZL glass (large $N$ value > 30 for the 25-µm waist diameter and the pumping in the anomalous dispersion regime) the coherence properties of the spectrum get worse as the peak power increases (see Fig. 5d).

With regard to the SC spatial content at the fiber output, we are able to characterize the modal content over the visible and near-IR range (see insets in Fig. 5a and 5c), our near-field imaging confirms that the fundamental mode was mainly excited for both uniform and tapered fibers. In the uniform fiber, no spectral broadening takes place so that it clearly provides a reference measurement in the nearly linear propagation regime. By contrast, in the tapered fiber, an ultra-broadband SC is generated through a highly nonlinear regime, nevertheless the SC light remains clearly supported by the fundamental mode. One can only notice a few side-perturbations in the picture that could be linked to intermodal couplings with non-circular symmetric higher-order modes. We also checked that SC extension can be optimized through the analysis of the output spatial profile (i.e., once the fundamental mode predominantly supports SC light).

3.4. High-power SC spectra in tapered glass rods

Subsequent experiments were conducted to raise the output SC average power in the different glass rods by increasing the repetition rate of our laser source step by step. Figure 6 illustrates the outcomes for the three different glasses and Table 1 summarizes the main experimental parameters. Notably, the spectrum presented for the GST fiber is the most optimized with a full mid-IR SC bandwidth of 140 THz, spanning from 2 to 15 µm and achieving an output average power of 93 mW for a 500-kHz repetition rate (see Fig.6(a)). Using suitable spectral filters, we also measured the average power beyond 9 µm, equal to 11 mW. We successfully operated at 1 MHz our tapered Ge-Se-Te rod with an output power of more than 170 mW. However, the SC spectrum did not extend up to 15 µm (see Fig.6(a)). We were able to repeat such SC generation at the 500-kHz and 1-MHz repetition rates in many tapered rods, but sometimes we also had irreversible damages, particularly on the fiber input facet. It is worth noting that the end facets of the Ge-Se-Te are cleaved manually with blades, which can cause some detrimental defects when operating at such high powers. Our measurements here demonstrate the combination of ultra-broad spectral broadening beyond 11 µm with robust performances at high repetition rates, supporting average power levels on the order of a hundred mW or more. Until now fiber-based SC spectra going far into the mid-infrared have been obtained with kHz-repetition-rate laser sources, which limit the output average power well below the mW level. Noteworthy examples of mid-IR SC with comparable average powers include the work by Petersen et al [17], achieving an average power of 35.4 mW at the fiber output for an SC spanning from 1 to 11.5 µm, and the SC reported by Hudson et al [18]. Figure 6(d) shows the state-of-the-art of mid-IR SC average power as a function of mid-IR spectral edge obtained at -20 dB for femtosecond pulses, but also for longer pulse durations (ps or ns). Only a cascaded fiber system in the long pulse pumping regime can overcome our performances in terms of average power (417 mW),[41] at the cost of spectral extension limited to 11 µm and an intrinsic low degree of coherence. In terms of power scaling, our scheme can be still optimized by means of anti-reflective coatings on fiber end facets. In the present case of 500-kHz repetition rate, we estimated our coupling efficiency to 55% of the delivered beam power available at 6 µm (by taking into account Fresnel losses for the Ge-Se-Te glass: 18% at each interface). We clearly overcome here the usual limitations associated with step-index or microstructured chalcogenide fibers, such as low damage threshold and power coupling restrictions due to small mode areas. Note that we made use of simple and easy-alignment long-focus coupling lenses.

| Glass | Pump wavelength (µm) | $n_{eff}$ ($\lambda_{pump}$) | $P_{crit}$ (MW) | Peak power (MW) | Average power (mW) | Repetition rate (kHz) | Waist diameter (µm) | SC bandwidth (µm) | SC bandwidth (THz) |
|---|---|---|---|---|---|---|---|---|---|
| GST | 6 | 2.49 | 0.26 | 0.72 | 93 | 500 | 40 | 1.9 - 14.9 | 140 |
| TZL | 2.5 | 1.97 | 1.24 | 1.85 | 237 | 500 | 40 | 0.6 - 5.0 | 440 |
| SiO$_2$ | 1.5 | 1.45 | 8.86 | 2.05 | 132 | 250 | 25 | 0.39 - 2.75 | 660 |

Table 1 – Parameters of SC generation in tapered glass rods depicted in Fig. 6.

For the TZL tapered rod, we successfully extended the SC spectrum up to 5 µm with an average power of 237 mW when operating at 500 kHz (see Fig.6(b)). In contrast to the GST rods, where an increase in the repetition rate led to damage issues on the input facet, working at higher rates with tellurite rods only revealed some irreversible damages in the waist section of tapered rods. The distinct resistance observed between chalcogenide and tellurite tapers can be linked to the different peak powers involved in the propagation, but also to their bond strength, a characteristic further illustrated by different glass transition temperatures (159°C for GST compared to 360°C for TZL[21,25]). For the silica fiber, the operational limit was constrained to a maximum repetition rate of 250 kHz resulting in a spectrum with an average power of 132 mW for a SC bandwidth of 660 THz (see Fig.6(c)). Attempts to increase the rate was not possible to prevent damage to the optics used in the experimental setup, which were designed to limit the energy of the pulse to µJ-level.

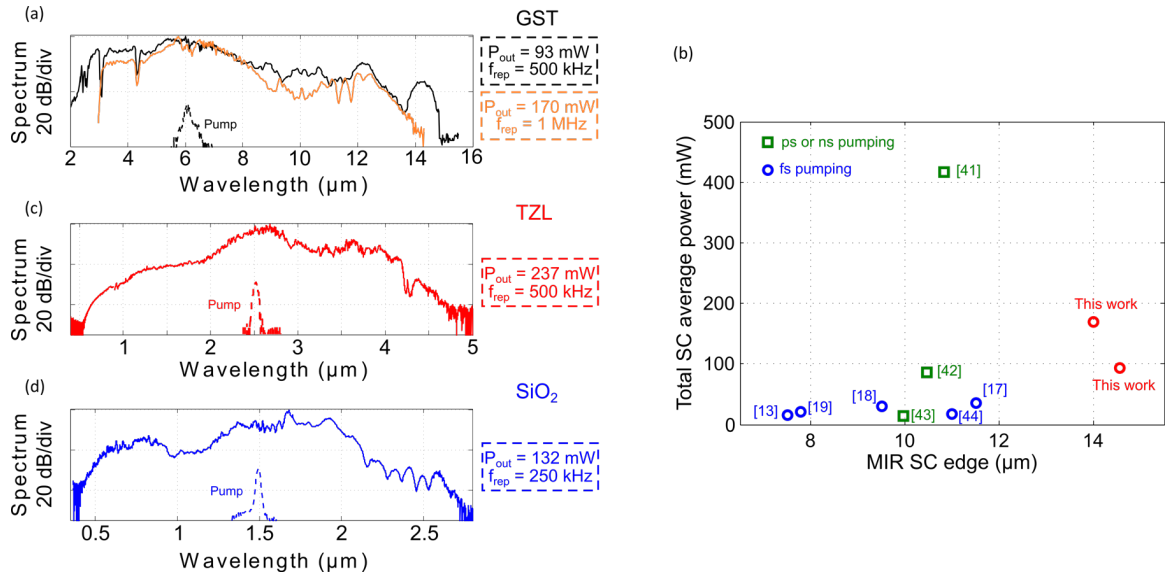

Figure 6- SC spectra obtained in tapered (a) GST fiber, (c) TZL fiber and (d) SiO$_2$ multimode fiber with maximum output average powers. (b Current state-of-the-art of mid-IR SC average power as a function of the long mid-IR spectral edge with direct pumping in the femtosecond regime and for cascaded systems with long pulses (ps or ns). The works cited are Refs [13,17–19,41–44].

## 4. Conclusion

In summary, we reported both experimental and numerical investigations of SC generation in few-cm long segments of several tapered glass rods made of single-glass materials and dedicated to distinct spectral windows. We successfully demonstrated the combination of ultra-broadband SC generation with high average powers by working with laser sources operating up to MHz and simple multimode fiber-based devices. Our simple post-processing of infrared glass rods offered new possibilities in terms of coupling efficiency, spectral coverage, and output power. As the most significant example, we unlocked the high-power regime for fiber-based supercontinuum sources beyond the 10 µm waveband, thus providing crucial average power for spectroscopy measurements at wavelengths where it is currently difficult to do so. Numerical simulations were used to confirm the major contribution of the fundamental mode in the ultrafast nonlinear dynamics, as well as the possible preservation of coherence features and spatial content. Our study then opens a new route to the power scaling of high-repetition-rate fiber supercontinuum sources over the full molecular fingerprint region. It is worth also mentioning that SC generation in multimode fibers can be achieved to facilitate nonlinear imaging with appropriate spatial resolution.[45]

**Funding Sources**

This work benefited from the facilities of the SMARTLIGHT platform funded by the Agence Nationale de la Recherche (EQUIPEX+ Contract No. ANR-21-ESRE-0040) and the Région Bourgogne Franche-Comté.

**Notes.** The authors declare no competing financial interest.

(21) Lemière, A.; Désévédavy, F.; Mathey, P.; Froidevaux, P.; Gadret, G.; Jules, J.-C.; Aquilina, C.; Kibler, B.; Béjot, P.; Billard, F.; Faucher, O.; Smektala, F. Mid-Infrared Supercontinuum Generation from 2 to 14 μm in Arsenic- and Antimony-Free Chalcogenide Glass Fibers. *J. Opt. Soc. Am. B* **2019**, *36* (2), A183. https://doi.org/10.1364/JOSAB.36.00A183.

(22) Sun, L.; Chen, F.; Xu, Y.; Huang, Y.; Liu, S.; Zhao, Z.; Wang, X.; Zhang, P.; Dai, S.; Zhang, X. Investigation of the Third-Order Nonlinear Property of Ge–Se–Te Glasses at Mid-Infrared. *Appl. Phys. A* **2016**, *122* (9), 816. https://doi.org/10.1007/s00339-016-0351-x.

(23) Couairon, A.; Mysyrowicz, A. Femtosecond Filamentation in Transparent Media. *Physics Reports* **2007**, *441* (2), 47–189. https://doi.org/10.1016/j.physrep.2006.12.005.

(24) Maldonado, A.; Evrard, M.; Serrano, E.; Crochetet, A.; Désévédavy, F.; Jules, J. C.; Gadret, G.; Brachais, C. H.; Strutynski, C.; Ledemi, Y.; Messaddeq, Y.; Smektala, F. TeO2-ZnO-La2O3 Tellurite Glass System Investigation for Mid-Infrared Robust Optical Fibers Manufacturing. *Journal of Alloys and Compounds* **2021**, *867*, 159042. https://doi.org/10.1016/j.jallcom.2021.159042.

(25) Evrard, M.; Combes, T.; Maldonado, A.; Désévédavy, F.; Gadret, G.; Strutynski, C.; Jules, J. C.; Brachais, C. H.; Smektala, F. TeO$_2$-ZnO-La$_2$O$_3$ Tellurite Glass Purification for Mid-Infrared Optical Fibers Manufacturing. *Opt. Mater. Express* **2022**, *12* (1), 136. https://doi.org/10.1364/OME.438588.

(26) Evrard, M.; Serrano, E.; Strutynski, C.; Désévédavy, F.; Gadret, G.; Jules, J. C.; Kibler, B.; Smektala, F. (INVITED)Dispersion-Shifted Tellurite Fibers for Nonlinear Frequency Conversion. *Optical Materials: X* **2022**, *15*, 100183. https://doi.org/10.1016/j.omx.2022.100183.

(27) Deroh, M.; Beugnot, J.-C.; Hammani, K.; Finot, C.; Fatome, J.; Smektala, F.; Maillotte, H.; Sylvestre, T.; Kibler, B. Comparative Analysis of Stimulated Brillouin Scattering at 2 Mm in Various Infrared Glass-Based Optical Fibers. *J. Opt. Soc. Am. B* **2020**, *37* (12), 3792. https://doi.org/10.1364/JOSAB.401252.

(28) Sunak, H. R. D.; Bastien, S. P. Refractive Index and Material Dispersion Interpolation of Doped Silica in the 0.6-1.8 Mu m Wavelength Region. *IEEE Photonics Technology Letters* **1989**, *1* (6), 142–145. https://doi.org/10.1109/68.36016.

(29) Tarnowski, K.; Majchrowska, S.; Béjot, P.; Kibler, B. Numerical Modelings of Ultrashort Pulse Propagation and Conical Emission in Multimode Optical Fibers. *J. Opt. Soc. Am. B* **2021**, *38* (3), 732. https://doi.org/10.1364/JOSAB.413050.

(30) Stefańska, K.; Béjot, P.; Tarnowski, K.; Kibler, B. Experimental Observation of the Spontaneous Emission of a Space–Time Wavepacket in a Multimode Optical Fiber. *ACS Photonics* **2023**, *10* (3), 727–732. https://doi.org/10.1021/acsphotonics.2c01863.

(31) Ravets, S.; Hoffman, J. E.; Kordell, P. R.; Wong-Campos, J. D.; Rolston, S. L.; Orozco, L. A. Intermodal Energy Transfer in a Tapered Optical Fiber: Optimizing Transmission. *J. Opt. Soc. Am. A* **2013**, *30* (11), 2361. https://doi.org/10.1364/JOSAA.30.002361.

(32) Kibler, B.; Béjot, P. Discretized Conical Waves in Multimode Optical Fibers. *Phys. Rev. Lett.* **2021**, *126* (2), 023902. https://doi.org/10.1103/PhysRevLett.126.023902.

(33) Stefańska, K.; Béjot, P.; Tarnowski, K.; Kibler, B. Experimental Observation of Spontaneous Emission of Space-Time Wavepacket in a Multimode Optical Fiber. **2022**. https://doi.org/10.48550/ARXIV.2208.06598.

(34) Moharram, A. H.; Hefni, M. A.; Abdel-Baset, A. M. Short and Intermediate Range Order of Ge20Se80−xTex Glasses. *Journal of Applied Physics* **2010**, *108* (7), 073505. https://doi.org/10.1063/1.3488907.

(35) Hollenbeck, D.; Cantrell, C. D. Multiple-Vibrational-Mode Model for Fiber-Optic Raman Gain Spectrum and Response Function. *J. Opt. Soc. Am. B* **2002**, *19* (12), 2886. https://doi.org/10.1364/JOSAB.19.002886.

(36) Yan, X.; Qin, G.; Liao, M.; Suzuki, T.; Ohishi, Y. Transient Raman Response and Soliton Self-Frequency Shift in Tellurite Microstructured Fiber. *Journal of Applied Physics* **2010**, *108* (12), 123110. https://doi.org/10.1063/1.3525595.

(37) Béjot, P. Multimodal Unidirectional Pulse Propagation Equation. *Phys. Rev. E* **2019**, *99* (3), 032217. https://doi.org/10.1103/PhysRevE.99.032217.

(38) Genty, G.; Coen, S.; Dudley, J. M. Fiber Supercontinuum Sources (Invited). *J. Opt. Soc. Am. B* **2007**, *24* (8), 1771. https://doi.org/10.1364/JOSAB.24.001771.

(39) Heidt, A. M.; Feehan, J. S.; Price, J. H. V.; Feurer, T. Limits of Coherent Supercontinuum Generation in Normal Dispersion Fibers. *J. Opt. Soc. Am. B, JOSAB* **2017**, *34* (4), 764–775. https://doi.org/10.1364/JOSAB.34.000764.

(40) Serrano, E.; Bailleul, D.; Désévédavy, F.; Gadret, G.; Mathey, P.; Béjot, P.; Nakatani, A.; Cheng, T.; Ohishi, Y.; Kibler, B.; Smektala, F. Multi-Octave Mid-Infrared Supercontinuum Generation in Tapered Chalcogenide-Glass Rods. *Opt. Lett., OL* **2023**, *48* (21), 5479–5482. https://doi.org/10.1364/OL.501036.

(41) Martinez, R. A.; Plant, G.; Guo, K.; Janiszewski, B.; Freeman, M. J.; Maynard, R. L.; Islam, M. N.; Terry, F. L.; Alvarez, O.; Chenard, F.; Bedford, R.; Gibson, R.; Ifarraguerri, A. I. Mid-Infrared Supercontinuum Generation from 1.6 to >11 μm Using Concatenated Step-Index Fluoride and Chalcogenide Fibers. *Opt. Lett., OL* **2018**, *43* (2), 296–299. https://doi.org/10.1364/OL.43.000296.

(42) Woyessa, G.; Kwarkye, K.; Dasa, M. K.; Petersen, C. R.; Sidharthan, R.; Chen, S.; Yoo, S.; Bang, O. Power Stable 1.5–10.5 μm Cascaded Mid-Infrared Supercontinuum Laser without Thulium Amplifier. *Opt. Lett., OL* **2021**, *46* (5), 1129–1132. https://doi.org/10.1364/OL.416123.

(43) Venck, S.; St-Hilaire, F.; Brilland, L.; Ghosh, A. N.; Chahal, R.; Caillaud, C.; Meneghetti, M.; Troles, J.; Joulain, F.; Cozic, S.; Poulain, S.; Huss, G.; Rochette, M.; Dudley, J. M.; Sylvestre, T. 2–10 μm Mid-Infrared Fiber-Based Supercontinuum Laser Source: Experiment and Simulation. *Laser & Photonics Reviews* **2020**, *14* (6), 2000011. https://doi.org/10.1002/lpor.202000011.
11